# Spectral Measures of Bipartivity in Complex Networks

Ernesto Estrada[1*] and Juan A. Rodríguez-Velázquez[2]

[1]*Complex Systems Research Group*, X-rays Unit, RIAIDT, Edificio CACTUS, University of Santiago de Compostela, 15706 Santiago de Compostela, Spain and [2]*Department of Mathematics*, University Carlos III de Madrid, 28911 Leganés (Madrid), Spain.

---

[*] To whom correspondence should be addressed. E-mail: estraad66@yahoo.com

We introduce a quantitative measure of network bipartivity as a proportion of even to total number of closed walks in the network. Spectral graph theory is used to quantify how close to bipartite a network is and the extent to which individual nodes and edges contribute to the global network bipartivity. It is shown that the bipartivity degree characterizes the network structure and can be related to the efficiency of semantic or communication networks, trophic interactions in food webs, construction principles in metabolic networks, or communities in social networks.

The study of complex networks has become an important area of multidisciplinary research involving physics, mathematics, biology, social sciences, informatics and other theoretical and applied sciences. The importance of this field resides in the existence of a unifying language to describe disparate real-world systems that are of great relevance in modern society, ranging from the Internet or power-grids to metabolic or protein interaction networks (PINs) [1]. Recently, "bipartivity" has been proposed as an important topological characteristic of complex networks [2]. A network (graph) $G = (V, E)$ is called *bipartite* if its vertex set $V$ can be partitioned into two subsets $V_1$ and $V_2$ such that all edges have one endpoint in $V_1$ and the other in $V_2$. There are numerous natural systems that are modelled as bipartite networks, such as reaction networks or "two-mode" networks, in which two disjointed sets of nodes are related by links representing the relationship between the elements of both classes [1,3]. For instance, they can represent authors that cite (or are cited by) papers, people that belong to institutions, cities that have certain services or voting results of delegates concerning certain proposals. Holme *et al*. [2] pointed out several areas for the potential application of a quantitative measure of bipartivity, such as network studies of sexually transmitted diseases, trade networks of buyers and sellers, "genealogical" networks of disease outbreak and food webs [2]. We introduce here a spectral measure of bipartivity for complex networks that is easy to compute, changes monotonically with changes in network bipartivity and allows the calculation of individual node contributions to global bipartivity, which is based on the concept of closed walks. A walk of length $r$ is a sequence of (not necessarily different) vertices $v_1, v_2, \cdots, v_r, v_{r+1}$ such that for each $i = 1, 2 \cdots, r$ there is a link from $v_i$ to $v_{i+1}$. A closed walk (CW) is a walk in which $v_{r+1} = v_1$. A CW is called odd (even) if $r$ is odd (even). A *cycle* is a CW in which all vertices in $v_1, v_2, \cdots, v_r, v_{r+1}$ are different. The theoretical motivation of our measure of bipartivity arose from the following result.

**Theorem**

A nontrivial graph (without self-loops) is bipartite if and only if it contains no odd closed-walks.

*Proof*

Let $G$ be a bipartite graph with partite sets $V_1$ and $V_2$. Suppose that $C: u = x_1, x_2,..., x_k = u$ is a CW in $G$ and that $u \in V_1$. Therefore, $u = x_1 \in V_1$, $x_2 \in V_2$, $x_3 \in V_1$, $x_4 \in V_2$, and so on. Hence, $C$ has even length. For the converse, it suffices to prove that every nontrivial connected graph $G$ without odd CWs is bipartite. Let $u, v \in V(G)$. We say that the subscripts of the vertex $x \in V(G)$ in the walk $u = x_0, x_1,..., x_k = v$ are $i_1, i_2,..., i_t$ if $x_{i_1} = x_{i_2} = \cdots = x_{i_t} = x$. Suppose that there are two subscripts, $i$ and $j$ ($i < j$), of the vertex $x$ in the walk $u = x_0, x_1,..., x_k = v$. Then, the subscripts $i$ and $j$ have the same parity: otherwise the CW $x = x_i, x_{i+1},..., x_{j-1}, x_j = x$ has odd length, which is a contradiction. Suppose that there are two walks from $u$ to $v$: $u = x_0, x_1,..., x_k = v$ and $u = y_0, y_1,..., y_r = v$. If the subscripts $k$ and $r$ have different parity, then the CW $u = x_0, x_1,..., x_{k-1}, v, y_{r-1},..., y_1, y_0 = u$ has odd length, which is a contradiction. Therefore, in all the walks from $u$ to $v$, all the subscripts of $v$ have the same parity; all are even or all are odd. Let $V_1$ denotes the subset of $V(G)$ consisting of $u$ and all vertices of $G$ with the property that all have subscripts even in the walks starting at $u$. Let $V_2 = V(G) - V_1$.

Now we prove that $G$ is bipartite whose partite sets are $V_1$ and $V_2$. Let $x, y \in V_1$, and suppose that $xy$ is an edge in $G$. In such case, neither $x$ nor $y$ is the vertex $u$. Therefore, there is at least a walk $u = x_0, x_1,..., x_{2k} = x$ of even length from $u$ to $x$, and there is at least a walk $u = y_0, y_1,..., y_{2r} = y$ of even length from $u$ to $y$. Hence, $G$ has at least a CW $u = x_0, x_1,..., x_{2k}, y_{2r}, y_{2r-1},..., y_1, y_0 = u$ of length odd, which is a contradiction to our hypothesis. The proof that no two vertices of $V_2$ are adjacent is similar.□

Now the following well-known result of graph theory [4] can be expressed as a corollary of the above result:

Corollary: *A nontrivial graph is bipartite if and only if it contains no odd cycles.*

Let $G = (V, E)$ be a network with $N$ nodes and having eigenvalues of the adjacency matrix $\sigma = (\lambda_1, \cdots, \lambda_N)$ [5]. The subgraph centralization measure has been defined as a weighted sum of all closed-walks (CWs) of length $l$ in the network, $\mu_l$, which is related to graph eigenvalues as follows [6]:

$$\langle SC \rangle = SC(G) = \frac{1}{N} \sum_{l=1}^{\infty} \frac{\mu_l}{l!} = \frac{1}{N} \sum_{j=1}^{N} e^{\lambda_j} \qquad (1)$$

This can be expressed as the sum of two contributions, one coming from odd and the other from even CWs:

$$\langle SC \rangle = \frac{1}{N} \sum_{j=1}^{N} \left( \cosh(\lambda_j) + \sinh(\lambda_j) \right) = \langle SC \rangle_{even} + \langle SC \rangle_{odd} \qquad (2)$$

If $G(V, E)$ is bipartite then: $\langle SC \rangle_{odd} = \frac{1}{N} \sum_{j=1}^{N} \sinh(\lambda_j) = 0$ because there are no odd CWs in the network and therefore:

$$\langle SC \rangle = \langle SC \rangle_{even} = \frac{1}{N} \sum_{j=1}^{N} \cosh(\lambda_j) \qquad (3)$$

Consequently, the proportion of even CWs to the total number of CWs is a measure of the network bipartivity:

$$\beta(G) = \frac{\langle SC \rangle_{even}}{\langle SC \rangle} = \frac{\langle SC \rangle_{even}}{\langle SC \rangle_{even} + \langle SC \rangle_{odd}} = \frac{\sum_{j=1}^{N} \cosh(\lambda_j)}{\sum_{j=1}^{N} e^{\lambda_j}} \qquad (4)$$

It is evident that $\beta(G) \leq 1$ and $\beta(G) = 1$ if, and only if, $G$ is bipartite, i.e., $\langle SC \rangle_{odd} = 0$. Furthermore, as $0 \leq \langle SC \rangle_{odd}$ and $\sinh(\lambda_j) \leq \cosh(\lambda_j)$, $\forall \lambda_i$, then $\frac{1}{2} < \beta(G)$ and $\frac{1}{2} < \beta(G) \leq 1$. The lower bound is reached for the least possible bipartite graph with $N$ nodes, which is the complete graph $K_N$. As the eigenvalues of $K_N$ are $N-1$ and $-1$ (with multiplicity $n-1$) [5], then $\beta(G) \to \frac{1}{2}$ when $N \to \infty$ in $K_N$. This lower bound coincides with that given by Holme *et al.* [2] in the $N \to \infty$ limit for their measures.

Despite the fact that both measures coincide in the extreme values, they show different values for the rest of networks. Consider for instance a process in which new edges are successively added to a bi-complete graph $K_{N1,N2}$, which has two disjointed sets of nodes $V_1$ and $V_2$ of cardinality $N_1$ and $N_2$, respectively. We will obtain the least bipartite graph $K_N$, $N = N_1 + N_2$, by joining together all of the $N_1$ nodes of $V_1$ and all of the $N_2$ nodes of $V_2$. The addition of one edge to $V_1$ ($V_2$) will introduce $N_2$ ($N_1$) triangles to the network. In the case of star graphs $K_{1,N2}$ the addition of one edge to $V_2$ introduces only one triangle to each graph, which make the graphs with larger $N_2$ more bipartite because the proportion of even to total CWs, $\beta(G)$, increases. In this case, $\beta(G)$ coincides with $b_1$, which is one minus the proportion of frustrated to total number of edges in the network. It can be seen that $b_1$ drops dramatically in $K_3$ (the triangle) despite it being very close to bipartite, indicating that the aforementioned proportion is the important aspect for bipartivity and not the number of frustrated edges to be removed to make the graph bipartite. Thus, both measures follow the same trend for $K_{1,N2}$ as shown in Figure 1 (a). However, both measures give different trends if we consider graphs of the type $K_{N1,N2}$ ($N_1 \neq 1$, $N_2 \neq 1$). In Figure 1 (b) we illustrate this situation by adding edges to the graph $K_{2,3}$. Here the addition of one edge to $V_2$ produces two triangles, while its addition to $V_1$ produces three. Thus the numbers of frustrated and total edges are the same, but the proportion of even to total CWs is not. As a consequence, $b_1$ gives the same value for these pairs of graphs. However, $\beta(G)$ shows that the graphs having the frustrated edge joining pairs of nodes at $V_2$ are more bipartite than those in which the frustrated edge is joining nodes at $V_1$. This result can be straightforwardly generalized to any kind of network.

A desired property for $\beta(G)$ is that it changes monotonically as the bipartivity of the graph changes. Let $G$ be a non-complete graph and let $e$ be an edge of the complement of $G$. Let $G+e$ be the graph obtained by adding the edge $e$ to $G$. In this situation, there exist real and non-negative numbers $a$ and $b$, such that $SC_{even}(G+e) = SC_{even}(G) + a$ and $SC_{odd}(G+e) = SC_{odd}(G) + b$. Notice that $a$ is the contribution of edge $e$ to $SC_{even}$ and $b$ is the contribution of edge $e$ to $SC_{odd}$. Thus, with the above notation, if $b \geq a$, then $\beta(G) \geq \beta(G+e)$. That is, as $\frac{a+b}{2} \geq a$ and $SC_{even}(G) \geq \frac{SC(G)}{2}$ then $(a+b) \cdot SC_{even}(G) \geq a \cdot SC(G)$. The addition of $SC_{even}(G) \cdot SC(G)$ to both terms and further reordering gives $SC_{even}(G) \cdot [SC(G) + a + b] \geq SC(G)[SC_{even}(G) + a]$ and, consequently:

$$\beta(G) = \frac{SC_{even}(G)}{SC(G)} \geq \frac{SC_{even}(G) + a}{SC(G) + a + b} = \beta(G+e) \tag{5}$$

which proves the monotony of the change for the spectral bipartivity measure as can be seen in the Figure 1 (b).

The contribution of node $i$ to network bipartivity, $\beta(i)$, can be obtained by using the subgraph centrality of node $i$ [6]:

$$SC(i) = \sum_{j=1}^{N} [v_j(i)]^2 e^{\lambda_j} \tag{6}$$

where $(v_1, v_2, ..., v_n)$ is an orthonormal basis of $R^N$ composed by eigenvectors of the adjacency matrix associated with the eigenvalues $\lambda_1, \lambda_2, ..., \lambda_N$, and $v_j(i)$ is the *i*th component of $v_j$. Hence, $\beta(i)$ is given by:

$$\beta(i) = \frac{\sum_{j=1}^{N}[v_j(i)]^2 \cosh(\lambda_j)}{\sum_{j=1}^{N}[v_j(i)]^2 e^{\lambda_j}} \qquad (7)$$

On the other hand, we can make use of our finding about the monotony of $\beta(G)$ to calculate the contribution of an edge $e$ to the network bipartivity $\beta(e)$. Let $e$ be an edge of the network $G$ and let $G-e$ be the network obtained by removing $e$ from $G$. Then, $\beta(e)$ is given by $\beta(e) = 1 - [\beta(G-e) - \beta(G)]$ [the formula $1 - [\cdots]$ is used to make these values follow the same trend as those of $\beta(i)$]. In Figure 1 (c) we illustrate the values of $\beta(i)$ and $\beta(e)$ for the nodes and links of the first two graphs of Figure 1(b).

Insert Figure 1 about here.

A total of 17 complex networks were studied, including two semantic networks, one based on Roget's Thesaurus of English (Roget) [7] and the other based on the Online Dictionary of Library and Information Science (ODLIS) [8] (two words are connected if one is used in the definition of the other); four social networks that include a scientific collaboration network in the field of computational geometry [9], inmates in prison [10], injecting drug users (IDUs) that have shared a needle in the last six months [11], and the friendship network between members of a karate club [12]; seven biological networks including the protein–protein interaction network (PINs) of yeast compiled by Bu *et al*. [13] on data obtained by von Mering *et al*. [14]; the direct transcriptional regulation between genes in yeast [15], and four food webs representing trophic relations in different ecosystems: Coachella Valley [16], Grassland [17], Ythan Estuary [18], and El Verde Rainforest [19]; five technological networks, one based on the airport transportation network in the US in 1997 [20], the Internet at the autonomous systems (AS) level as from April 1998 [21], and three electronic sequential logic circuits parsed from the ISCAS89 benchmark set, where nodes represent logic gates and flip-flops [22].

The results of calculations are given in Table 1. There are four networks in which the low values of $\beta(G)$ are indicative of the efficient construction of these networks. These are the two semantic networks, the transportation network of USA airports and the Internet at AS. In dictionaries, like Roget and ODLIS, all individual entries must be bootstrapped from other entries in a self-referential way [23], which immediately precludes bipartivity from these semantic networks. In transportation or communication networks, a significant degree of bipartivity is translated into a low efficiency in travelling between the nodes located in the same disjointed set, which makes the network inefficient. However, bipartivity can also be a desired property in technological networks, as demonstrated by the high bipartivity observed for the three electronic circuits studied.

Insert Table 1 about here.

The four social networks analyzed show very different bipartivity values. While the karate club and prison networks reveal certain bipartivity, the IDUs and the collaboration network show values of $\beta(G) = 0.5$. The bipartivity observed for the Karate Club network can be rationalized by the fact that there are two main disjointed *hubs* in the network: the club's instructor and the club's president. The rest of the nodes, which form the other set, show a low average degree (3.84), indicating that there is not a high number of links between them. At the other extreme is the collaboration network, which has a value of $\beta(G) = 0.5$. This network consists of clusters of fully connected nodes, formed by coauthors of a particular paper, which are interconnected and make the network non-bipartite.

Holme *et al*. [2] considered the study of food webs as a potential area for the application of bipartivity measures. This supposition is based on the idea that the simplest picture of a food web can be represented as different "trophic" levels where species in one level predate species located at the level below, producing networks with a high degree of bipartivity. This appears to be the case for Grassland and Stony stream, both of which have a significant degree of bipartivity. Grassland represents a good example of a network where trophic levels are responsible for the bipartivity observed. The trophic relations observed are only inter-classes and forms an almost bipartite graph (graphic not shown). The situation is quite different for the other two food webs, which show a very low degree of bipartivity as a consequence of their larger number of trophic relations between species'. Similar results ($\beta \approx 0.5$) are obtained for other food webs, such as Little Rock, Scotch Broom and Ythan Estuary with parasites (data not shown). We therefore believe that bipartivity is not a general characteristic of ecological systems, despite the fact that food webs with a pronounced degree of bipartivity can be found as a consequence of the trophic relations between classes.

The metabolic network of yeast is the most bipartite network of all those studied here. This finding can be explained by the construction of the network, which is based on two set of nodes − one representing regulating genes and the other representing regulated genes − with connections between both sets. In contrast, the PIN of yeast shows a low degree of bipartivity despite its low clustering, which indicates that odd-cycles larger than triangles play an important role in the interactions between proteins in this organism.

The utility of the local bipartivity index $\beta(i)$ lies in the possibility of identifying those nodes and links that contribute significantly to the bipartivity in a network. Removing them will significantly affect the bipartivity degree of the whole network. For instance, removing the node with the lowest contribution to $\beta(i)$ in Grassland increases its bipartivity from $\beta(G) = 0.734$ to $\beta(G) = 0.794$ and this example can reach $\beta(G) = 0.863$ by removing the three nodes with the lowest contribution to $\beta(i)$. It is possible to find numerous practical applications for the detection of node/link bipartivity in real-world networks. For instance, this approach can be applied in the field of sexually transmitted diseases, where nodes with different contributions to the network bipartivity can play different roles in the transmission of such diseases [2].

Network bipartivity is a topological characteristic that can not be accounted for by other structural measures, such as clustering coefficients (see Table 1). On the other hand, the physical consequences of network bipartivity depend on the particular network and processes which are studied and cannot be generalized even for networks in the same fields. While social networks of friendships are expected to be non-bipartite due to the propensity for two of one's friends to also be friends of each other, in sexual networks bipartivity can arise from heterosexual contacts. A similar situation occurs in food webs, where in some systems the trophic relations between species in different trophic levels can introduce bipartivity to the network – a situation that does not occur if the species are in the same trophic level. The exception appears to be communication/information systems, in which the lack of bipartivity represents a measure of efficiency in the network construction.

We would like to thank J. A. Dunne, R. Milo, U. Alon, J. Moody and V. Batagelj for providing datasets. EE thanks the "Ramón y Cajal" program, Spain, for partial financial support.

FIG. 1. Some graphs used in the discussion of global and local spectral bipartivity measures. (a) Graphs obtained by the addition of one frustrated edge (dotted line) to stars $K_{1,N_2}$. Only one triangle is introduced to each graph after the addition of the new links. Thus the proportion of even to total CWs (even + odd), $\beta(G)$, increases as $N_2$ increases and the networks with larger $N_2$ are more bipartite. The proportion of frustrated to total number of edges decreases as $N_2$ increases and $b_1$ follows the same trend that $\beta(G)$. (b) Values of network bipartivity, $b_1$ and $\beta(G)$, for the graphs produced by successive addition of frustrated edges to $K_{2,3}$. It is observed that $\beta(G)$ shows a monotonically decreasing trend from the bipartite graph to the complete graph. $b_1$ do not differentiate between pairs of graphs having the same number of frustrated edges despite they show different proportions of even to total number of CWs. (c) Values of $\beta(i)$ and $\beta(e)$ for the first two graphs in (b). The lowest values of these measures indicate those nodes/links which contribute more to the non-bipartivity of the graph. Despite some links with values different from one are not frustrated edges their removing reduce the number of odd CWs, e.g., triangles, increasing the bipartivity of the network.

Figure 1

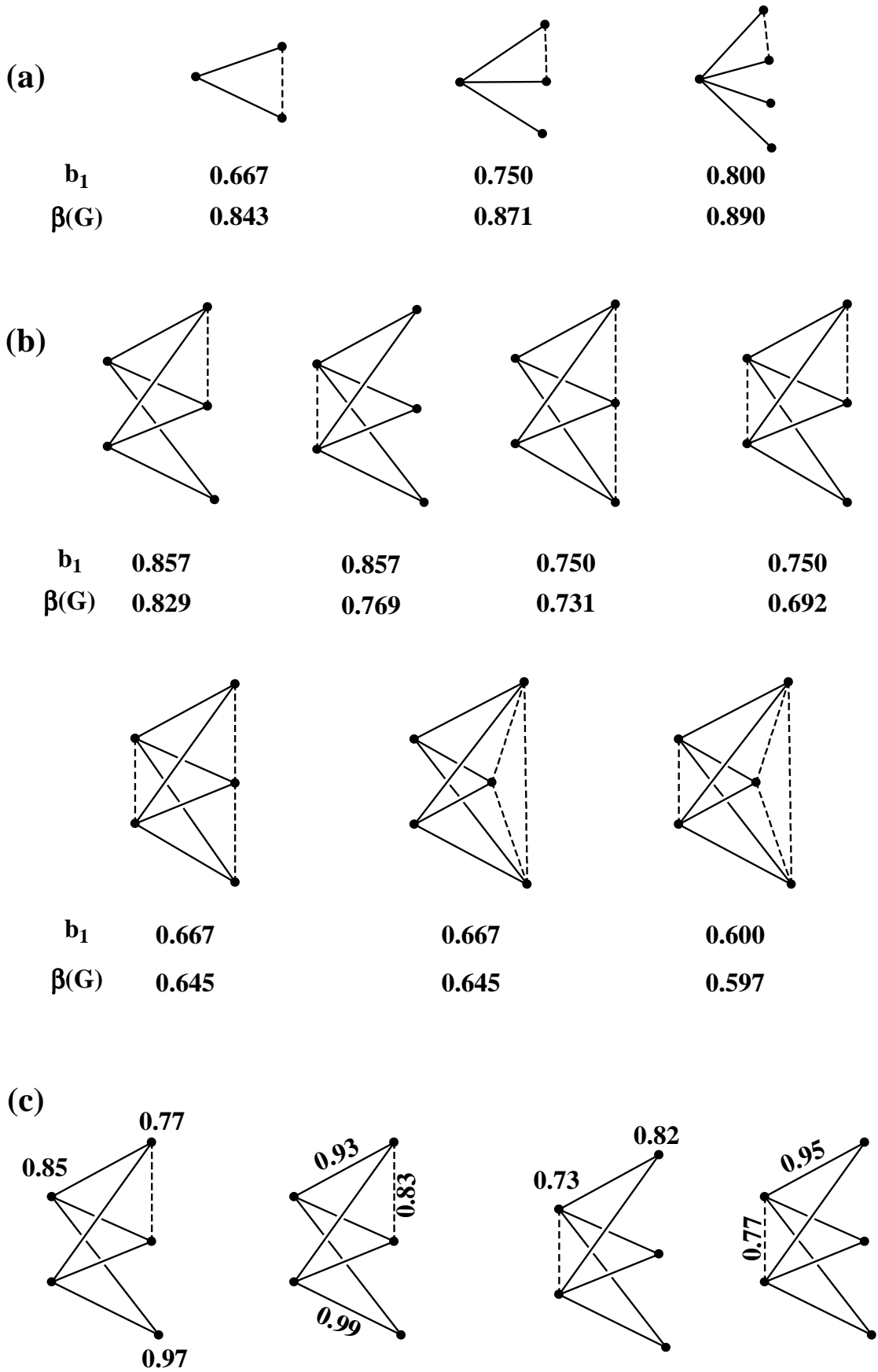

Table 1. Values of the spectral network bipartivity measure, $\beta(G)$, for complex networks of different types and sizes. The number of nodes (N) and edges (E) are given along with clustering coefficients. The correlation coefficients ($R^2$) of the linear regression between bipartivity and clustering coefficients are also given in order to show their linear independence.

| Type | Network | N | E | $\beta(G)$ | $C^{(1)}(G)$ | $C^{(2)}(G)$ |
|---|---|---|---|---|---|---|
| Information | Roget | 994 | 3640 | 0.529 | 0.162 | 0.134 |
|  | ODLIS | 2898 | 16376 | 0.500 | 0.351 | 0.056 |
| Social | Karate Club | 34 | 78 | 0.597 | 0.588 | 0.256 |
|  | Prison | 67 | 142 | 0.698 | 0.330 | 0.288 |
|  | Drugs | 616 | 2012 | 0.500 | 0.722 | 0.368 |
|  | Geom | 3621 | 9461 | 0.500 | 0.679 | 0.219 |
| Biological | Coachella | 30 | 241 | 0.500 | 0.707 | 0.697 |
|  | Grassland | 75 | 113 | 0.743 | 0.497 | 0.174 |
|  | Stony stream | 112 | 830 | 0.815 | 0.076 | 0.020 |
|  | El Verde | 156 | 1439 | 0.500 | 0.231 | 0.232 |
|  | Trans-yeast | 662 | 1062 | 0.960 | 0.092 | 0.016 |
|  | PIN-yeast | 2224 | 6608 | 0.500 | 0.201 | 0.102 |
| Technological | USAir97 | 332 | 2126 | 0.500 | 0.749 | 0.396 |
|  | Internet-1998 | 3522 | 6324 | 0.502 | 0.340 | 0.014 |
|  | Electronic1 | 122 | 189 | 0.948 | 0.064 | 0.344 |
|  | Electronic2 | 252 | 399 | 0.950 | 0.060 | 0.310 |
|  | Electronic3 | 512 | 819 | 0.952 | 0.058 | 0.290 |
| $R^2$ |  |  |  |  | 0.450 | 0.100 |

$C^{(1)}(G)$ as defined by Watt and Strogatz and $C^{(2)}(G)$ defined as 3 times the number of triangles divided by the number of connected triples in the network (see [1] for definitions).